# Study of spin fluctuations in $Ni_{3\pm x}Al_{1\mp x}$ using point contact Andreev reflection spectroscopy

Sourin Mukhopadhyay, Sudesh Kumar Dhar, Pratap Raychaudhuri[a]

Department of Condensed Matter Physics and Materials Science, Tata Institute of Fundamental Research, Homi Bhabha Rd., Colaba, Mumbai 400005, India.

**Abstract:**

We report point contact Andreev reflection (PCAR) spectroscopy studies on $Ni_{3\pm x}Al_{1\mp x}$ with composition range (24, 26 and 27 at. % of Al in the solid solution) spanning the ferromagnetic to paramagnetic phase boundary. PCAR studies performed using Nb tip as counter electrode reveal that the superconducting quasiparticle lifetime ($\tau$) and superconducting energy gap ($\Delta$) decreases with increasing spin fluctuation in the normal metal electrode. Our study reveals that PCAR could be a useful probe to study spin fluctuations in systems which are on the verge of a magnetic instability.

[a] Electronic-mail: pratap@tifr.res.in

Point contact spectroscopy[1] has been used for many years to investigate the static and dynamic spin dependent properties of magnetic systems at the Fermi level ($E_F$), *viz.* electron-magnon interaction[2], Kondo scattering[3] *etc*. In recent years a variant of this technique, namely, point contact Andreev reflection (PCAR) has emerged as a popular technique to measure the spin polarization at $E_F$ in itinerant ferromagnets[4]. In this technique a ballistic contact is established between a sharp superconducting tip and a normal metal electrode. In such a contact the electrical transport is governed by the process of Andreev reflection where an electron incident from the normal metal electrode on the normal-metal/superconductor interface is reflected as a hole in the opposite spin band and a Cooper pair propagates in the superconductor. This causes a doubling of the differential conductance of the junction for bias voltages lower than the superconducting energy gap. In a ferromagnet the Andreev reflection is suppressed due to the unequal density of states of up and down spins. The suppression in PCAR conductance spectra gives a measure of transport spin polarization[5] ($P_t$) at the Fermi level. In recent papers[6] we have shown that in addition to transport spin polarization PCAR can also be utilized to detect spin fluctuations in systems that are close to a magnetic instability. The proximity effect between a spin fluctuating magnetic electrode and the superconductor causes a decrease in the superconducting quasiparticle lifetime ($\tau$), thereby broadening[7,8] the Bardeen Cooper Schrieffer (BCS)[9] density of states. The broadened density of states is given[10] by, $N(E) = Re\left(\frac{E+i\Gamma}{\sqrt{(E+i\Gamma)^2-\Delta^2}}\right)$, where the broadening parameter ($\Gamma$) is given by, $\Gamma = \frac{\hbar}{\tau}$. From the magnitude of $\Gamma$, we get an estimate about the degree of spin fluctuation present in the system[6].

In this paper we report PCAR spectroscopy on the itinerant ferromagnet $Ni_{3\pm x}Al_{1\mp x}$. $Ni_{3\pm x}Al_{1\mp x}$ has been widely studied[11,12,13] due to its unusual compositional phase boundary around 24-27 at. % Al. Around this composition, the ground state of this system transforms from a ferromagnet to a paramagnet with large spin fluctuations. The parent alloy $Ni_3Al$ and its derivatives, in the composition range 23-27.5 at. % of Al, crystallize with a simple cubic structure of the form $Cu_3Au$. In the composition range 23-26 at.% of Al, $Ni_{3\pm x}Al_{1\mp x}$ exhibits ferromagnetism with moderate Curie temperature in the temperature range ~ 40-80 K; with increased Al concentrations above ~26 at. % of Al it becomes a paramagnet with large spin fluctuation. In this paper, we study the effect of spin fluctuation on the PCAR spectra of $Ni_{3\pm x}Al_{1\mp x}$ with composition range spanning the ferromagnet to paramagnet compositional phase boundary.

$Ni_{3\pm x}Al_{1\mp x}$ samples are prepared through arc melting. Detailed compositional analysis was carried out using electron probe microanalysis. Details of sample preparation and characterization have been described elsewhere[11]. To characterize the sample magnetization measurements are carried out on all the samples using a Vibrating Sample Magnetometer in the temperature range 2K-300K. Magnetoresistance measurements are carried out up to a field of 12T using a Quantum design Physical Property Measurement System. Point contact measurements in concerned samples are performed with a superconducting Nb tip in a liquid He-4 continuous flow cryostat using standard four-probe lock-in technique.

We have chosen three alloys namely $Ni_{76}Al_{24}$, $Ni_{74}Al_{26}$ and $Ni_{73}Al_{27}$. Fig. 1.(a-c) shows the temperature dependence of magnetic susceptibility M(T) down to 2K. $Ni_{76}Al_{24}$ shows a clear ferromagnetic transition around 72K (determined from the maximum in the double derivative of the M-T curve). $Ni_{74}Al_{26}$ also shows a ferromagnetic transition around 60K, but the moment is considerably smaller than $Ni_{76}Al_{24}$ and the magnetization does not saturate down to the lowest temperature indicating the presence of spin fluctuations even below $T_C$. $Ni_{73}Al_{27}$ does not exhibit any ordering down to the lowest temperature. However, unlike conventional Pauli paramagnets the magnetization is temperature independent showing a Curie-Weiss like behavior. This shows the presence of large spin fluctuation in the paramagnetic state. To confirm the presence of spin fluctuations, we have measured the temperature variation of magnetoresistance (MR=$\Delta\rho/\rho$=[$\rho$(H,T)-$\rho$(0,T)]/$\rho$(0,T)) at H=12 T as shown in Fig.(1d) for the two extreme compositions, namely, $Ni_{76}Al_{24}$ and $Ni_{73}Al_{27}$. For $Ni_{76}Al_{24}$, at $T<T_{Curie}$, the MR is low and increases with temperature and shows a maximum close to the Curie temperature. This can be understood since presence of strong ferromagnetic spin fluctuation around $T_{Curie}$ increases the resistance $\Delta\rho$. On the other hand, in $Ni_{73}Al_{27}$, the MR keeps increasing with decreasing temperature showing the presence of strong spin fluctuations down to the lowest temperature.

Fig. 2 (a-c) shows representative PCAR spectra for $Ni_{76}Al_2$, $Ni_{74}Al_{26}$ and $Ni_{73}Al_{27}$ respectively. These spectra are analyzed within the modified Blonder-Tinkham-Klapwijk (mBTK) formalism[14] using $\Delta$, $P_t$, $\Gamma$ and an effective barrier potential ($Z$) as fitting parameters. The barrier potential takes into account[15] any possible barrier at the interface arising from both possible oxide layers as well as the mismatch between Fermi velocities of the normal metal and superconducting electrode. This potential is modeled[16] as a delta function, $V(x) = V_0\delta(x)$ and is parameterized by a dimensionless quantity $Z= V_0/\hbar v_F$ where $v_F$ is the Fermi velocity. The values of $P_t$, $\Delta$ and $\Gamma$ extracted from the mBTK fits is shown as a function of at. % of Al in Fig. 2(d). With increasing Al, the spin polarization decreases and as expected is zero for the paramagnetic compound $Ni_{73}Al_{27}$. The broadening parameter $\Gamma$ is zero for the ferromagnetic $Ni_{76}Al_{24}$ sample and increases gradually with Al content. $\Delta$ on the other hand decreases with increasing Al content as one goes from the ferromagnetic to the spin fluctuating regime. This is consistent with our earlier results on $NdNi_5$, where we showed[6] that the proximity of the superconductor to large spin fluctuations drastically decreases the superconducting quasiparticle lifetime and the superconducting energy gap. To cross check the uniqueness of the fit and the influence of individual parameters on it, the PCAR data for $Ni_{73}Al_{27}$-Nb contact is also fitted assuming $\Gamma$=0 and a finite $P_t$ [dashed line in Fig. 2(c)] The fit is considerably poorer particularly at voltage values above the superconducting energy gap. This indicates that the large decrease in $\tau$ (increase in $\Gamma$) as seen in $Ni_{73}Al_{27}$-Nb contacts can be related to the increased spin fluctuations in the same, which is developing with increased Al concentration.

To further explore the effect of spin fluctuations on $\Gamma$, we have done detailed PCAR studies on the two extreme compositions of this compound *viz.* the ferromagnetic $Ni_{76}Al_{24}$ and the spin fluctuating $Ni_{73}Al_{27}$. Different PCAR spectra are recorded by engaging the Nb tip several times on them at different places. These spectra correspond to the same sample tip combinations but have statistically different values of Z. Fig. 3(a) shows the variation of $P_t$ with the barrier potential $Z$ for the ferromagnetic $Ni_{76}Al_{24}$. As has been shown before[17,18], the value of $P_t$ decreases with increasing Z due the presence of magnetic dead layer in the F-S interface, where spin flip scattering depolarises the electron, thereby decreasing the value of the transport spin polarization. Similar studies on $Ni_{73}Al_{27}$ reveal a systematic variation between $Z$ and $\Gamma$. With increasing $Z$, $\Gamma$ decreases [Fig. 3(b)]. This is expected since a larger barrier parameter at the interface implies that the two electrodes are less strongly coupled to each other. Therefore, the influence of spin fluctuation would be less on the superconducting electrode. Consequently, there is also an inverse correlation between $\Delta$ and $\Gamma$ extracted for contact with different $Z$ [*inset* figure 3(b)]. The smaller the lifetime of the quasiparticle, the smaller is $\Delta$ for the superconductor. This inverse correlation between $\Delta$ and $\Gamma$ provides a valuable consistency check of intrinsic nature of proximity effect between a superconductor and a spin fluctuating metal.

In summary, we have investigated the spin fluctuations in the itinerant ferromagnet $Ni_{3\pm x}Al_{1\mp x}$ using transport, magnetization and PCAR spectroscopy where the ground state evolves from a ferromagnet to a spin fluctuating paramagnet with increase in Al. Our study shows that PCAR spectroscopy can detect the signature of spin fluctuations through a decrease of the superconducting quasiparticle lifetime and superconducting energy gap due to proximity effect. The central observation of this paper is that while a static moment has negligible effect on the superconducting quasiparticle lifetime and superconducting energy gap extracted from PCAR spectra, spin fluctuations decreases both the lifetime of the quasiparticle and the superconducting energy gap. This study shows that PCAR can be a valuable tool to explore spin fluctuations in itinerant systems and its effect on superconductivity.

The authors would like to thank A. Thamizhavel and D.A. Joshi for helpful discussions, Subash Pai, Manish Ghag and Ruta N. Kulkarni for technical help, and Kavita Bajaj for her participation in early parts of these experiments.

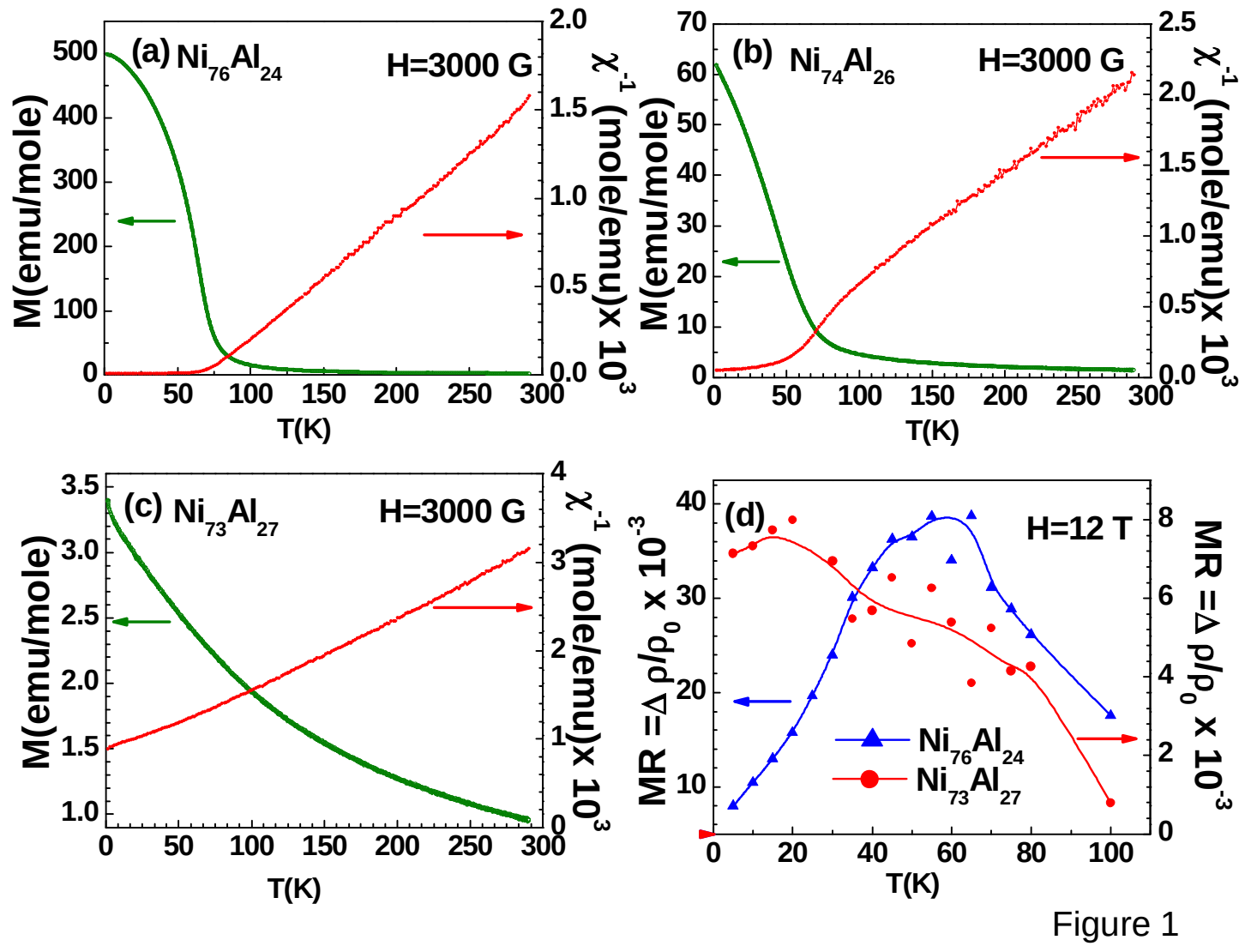

Fig. 1 (Color online) (a-c) shows the temperature (T) variation of magnetization (M) and inverse susceptibility ($\chi^{-1}$) for $Ni_{76}Al_{24}$, $Ni_{74}Al_{26}$ and $Ni_{73}Al_{27}$ respectively. (d) Magnetoresistance (MR) as a function of temperature (T) for $Ni_{76}Al_{24}$ and $Ni_{73}Al_{27}$ respectively. Solid lines are a guide to eye.

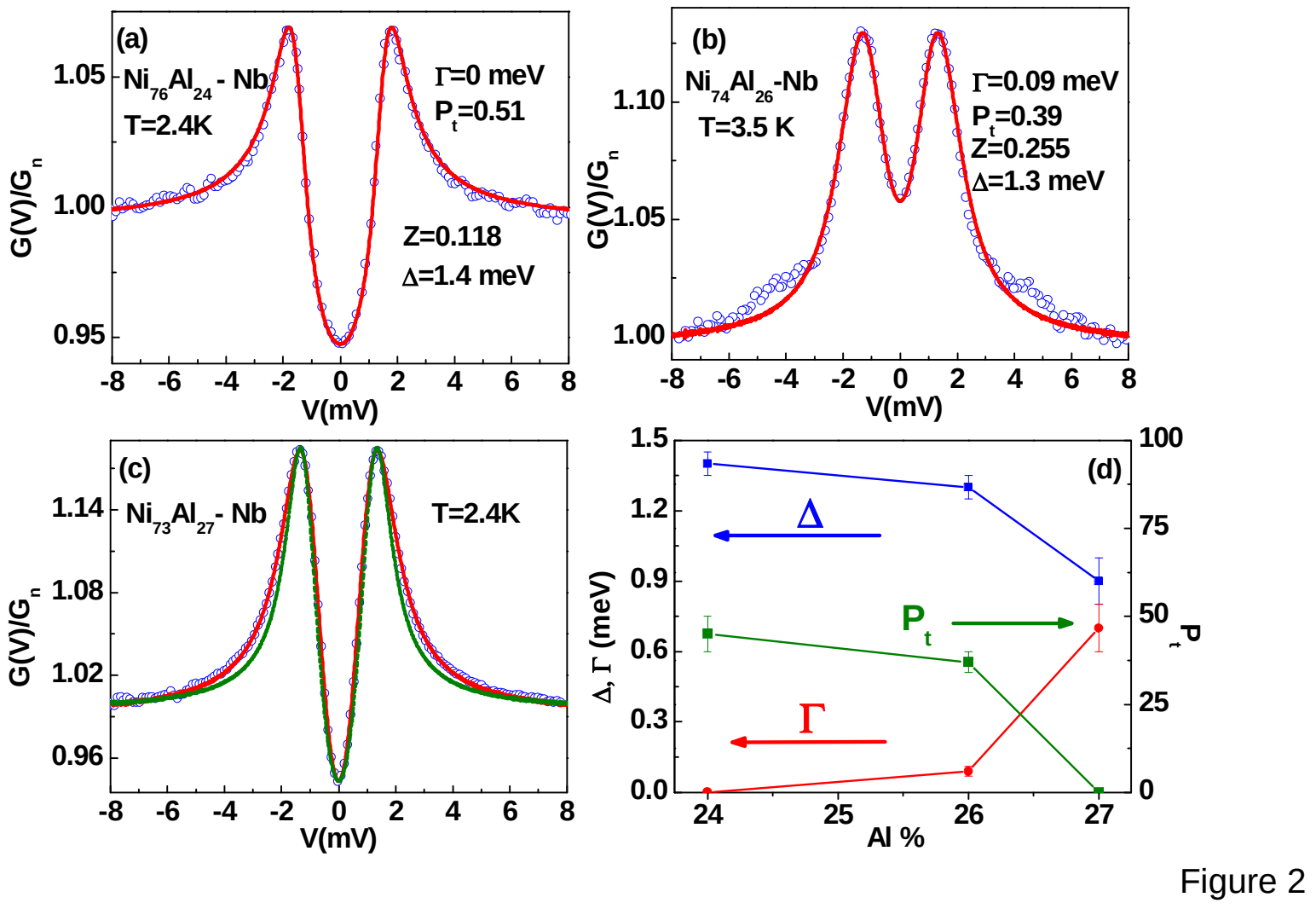

Fig. 2 (Color online) (a-b) PCAR spectra for $Ni_{76}Al_{24}$ and $Ni_{74}Al_{26}$ with mBTK fits (solid line). (c) PCAR spectra $Ni_{73}Al_{27}$; solid (red) is the mBTK fit to data with best fit parameters $P_t$=0, $\Gamma$=0.44 meV, $\Delta$=1.18 meV and $Z$=0.72. Dashed (green) is simulated spectra for the same with $\Gamma$=0, $\Delta$=1.28 meV and $Z$=0.32 and $P_t$~39%.(d) Variation of $\Delta$, $\Gamma$ and $P_t$ with Al concentration.

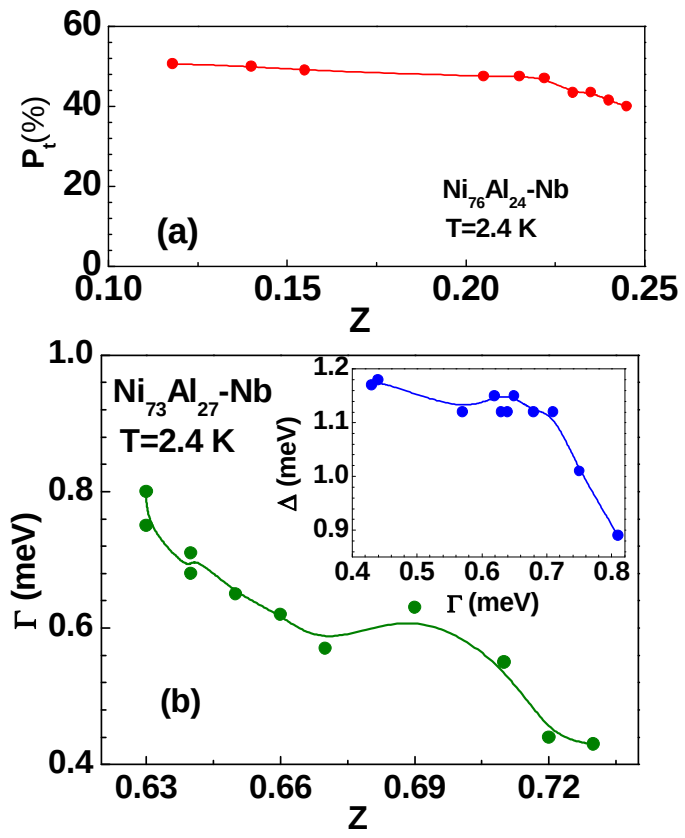

Figure 3

Fig. 3(a) (Color online)The variation of $P_t$ with $Z$ at 2.4 K for different $Ni_{76}Al_{24}$-Nb contacts, (b) Variation of the extracted value of $\Gamma$ with $Z$ at 2.4 K for different contacts. Inset shows variation of $\Delta$ with $\Gamma$. Solid lines are a guide to eye.